# A zero-test for D-algebraic transseries*

BY SHAOSHI CHEN[abC], HANQIAN FANG[ad], JORIS VAN DER HOEVEN[efG]

*a*. KLMM, Academy of Mathematics and Systems Science
Chinese Academy of Sciences,
Beijing 100190, China

*e*. LIX, CNRS, Institut Polytechnique de Paris
Bâtiment Alan Turing, CS35003
1, rue Honoré d'Estienne d'Orves
91120 Palaiseau, France

*b*. *Email:* `schen@amss.ac.cn`
*d*. *Email:* `fanghanqian22@mails.ucas.ac.cn`
*f*. *Email:* `vdhoeven@lix.polytechnique.fr`

*draft version of February 1, 2026*

**Abstract**

Consider formal power series $f_1, \ldots, f_k \in \mathbb{Q}[[z]]$ that are defined as the solutions of a system of polynomial differential equations together with a sufficient number of initial conditions. Given $P \in \mathbb{Q}[F_1, \ldots, F_k]$, several algorithms have been proposed in order to test whether $P(f_1, \ldots, f_k) = 0$. In this paper, we present such an algorithm for the case where $f_1, \ldots, f_k$ are so-called transseries instead of power series.

## 1 Introduction

Standard mathematical notation exists for many special functions such as exp, log, sin, erf, Si, $\wp$, $\zeta$, etc. But how to decide whether an expression like $\mathrm{e}^{x+y} - \mathrm{e}^x \mathrm{e}^y$ actually represents the zero function?

One popular approach is to rely on differential algebra [25, 20, 2], by defining exp as a symbolic solution of the equation $f' = f$. However, this only allows one to define exp up to a multiplicative constant, which is insufficient to conclude that $\mathrm{e}^{x+y} = \mathrm{e}^x \mathrm{e}^y$.

Another approach is to define $f = \exp$ as the unique solution of $f' = f$ with $f(0) = 1$. This can be made more precise by restricting our attention to D-algebraic power series. Let $\mathbb{K}$ be an effective field. We say that $f \in \mathbb{K}[[z]]$ is *D-algebraic* if it satisfies an equation $P(f, f', \ldots, f^{(r)}) = 0$ for some differential polynomial $P \in \mathbb{K}[F, F', \ldots, F^{(r)}] \setminus \mathbb{K}$. In that case, $f$ is actually uniquely determined by $P$ and a sufficient number of initial conditions. Given D-algebraic series $f_1, \ldots, f_k$ and a differential polynomial $Q$ in $F_1, \ldots, F_k$, the *zero-test problem* now consists of deciding whether $Q(f_1, \ldots, f_k) = 0$. Many solutions have been proposed for this problem [24, 5, 6, 19, 26, 27, 22, 17, 15] and we refer to [12] for a brief overview of existing approaches.

*C*. S. Chen and H. Fang were partially supported by National Key R&D Programs of China (number 2023YFA1009401), the NSFC grants (number 12271511), and the Strategic Priority Research Program of the Chinese Academy of Sciences (number XDB05102). All authors were also supported by the International Partnership Program of Chinese Academy of Sciences (Grant number 167GJHZ2023001FN).

*G*. J. van der Hoeven has been supported by an ERC-2023-ADG grant for the ODELIX project (number 101142171).

*. This article has been written using GNU $\mathrm{T_EX_{MACS}}$ [16].

However, one problem with ordinary power series is that they do not form a field. So-called *transseries* are a far reaching generalization of power series, by closing off $\mathbb{K}[[z]]$ under exponentiation, logarithms, and infinite summation [4, 8, 7, 14]. An example of a transseries at infinity ($x \to \infty$) is

$$\mathrm{e}^{\mathrm{e}^x + \frac{\mathrm{e}^x}{x} + \frac{\mathrm{e}^x}{x^2} + \cdots} + 3\,\mathrm{e}^{x \log x} + \sqrt{2} + \frac{1}{x} + \frac{2}{x^2} + \frac{6}{x^3} + \frac{24}{x^4} + \cdots. \tag{1}$$

The transseries with real coefficients form a field $\mathbb{T}$ that is closed under differentiation, composition and the resolution of many differential equations. A transseries $f$ is again said to be *D-algebraic* if $P(f, f', \ldots, f^{(r)}) = 0$ for some $P \in \mathbb{K}[F, F', \ldots, F^{(r)}] \setminus \mathbb{K}$. For example, the transseries (1) is D-algebraic.

Given D-algebraic transseries $f_1, \ldots, f_k$ and a differential polynomial $Q$ in $F_1, \ldots, F_k$, our main result is an algorithm for deciding whether $Q(f_1, \ldots, f_k) = 0$. In fact, we reduce this problem to the case $k = 1$, but at the same time generalize it to the case where the coefficients of $Q$ and the differential polynomial $P$ with $P(f_1) = 0$ are taken in an effective differential subfield of $\mathbb{T}$ for which we already have a zero-test.

A substantial part of the paper is devoted to recalling required theoretical results about transseries from [14, 11, 1]. From an effective point of view, computations with transseries are most conveniently carried out with respect to a so-called *transbasis* $\mathfrak{B} = (\mathfrak{b}_1, \ldots, \mathfrak{b}_n)$. This allows one to consider general transseries as power series in $\mathfrak{b}_n^{-1}$ with real exponents, and whose coefficients can recursively be expanded with respect to the transbasis $(\mathfrak{b}_1, \ldots, \mathfrak{b}_{n-1})$. For instance, for $x \to \infty$, the transseries $1/(1 - x^{-1} - \mathrm{e}^x)$ is a series in $\mathrm{e}^{-x}$ with coefficients in $\mathbb{R}[[x^{-1}]]$. Along with our survey, we present a precise framework, much in the vein of [23, 10].

Having carried out the necessary preparations, our main algorithm turns out to be a natural extension of the zero-test from [15] for formal power series. Using a theorem from [11], we were also able to further sharpen the bound on the required expansion order. As an illustration of our algorithm, we will first define the Lambert function as the unique transseries solution to a suitable asymptotic differential equation and then verify that it satisfies the equation $W(x)\,\mathrm{e}^{W(x)} = x$.

# 2 Generalized power series

## 2.1 Grid-based series

Let $\Gamma$ be a totally ordered abelian group. A subset $S \subseteq \Gamma$ is said to be *grid-based* if there exist $\alpha_1, \ldots, \alpha_k \in \Gamma^{>} := \{\gamma \in \Gamma : \gamma > 0\}$ and $\beta \in \Gamma$ with

$$S \;\subseteq\; \mathbb{N}\,\alpha_1 + \cdots + \mathbb{N}\,\alpha_k + \beta \;:=\; \{i_1\,\alpha_1 + \cdots + i_k\,\alpha_k + \beta : i_1, \ldots, i_k \in \mathbb{N}\}.$$

Given a field $\mathbb{K}$, consider a formal power series $f = \sum_{\alpha \in \Gamma} f_\alpha z^\alpha$ with $f_\alpha \in \mathbb{K}$ and such that the support $\operatorname{supp} f := \{\alpha \in \mathbb{R} : f_\alpha \neq 0\}$ is grid-based. Then we call $f$ a *grid-based power series* (with coefficients in $\mathbb{K}$ and exponents in $\Gamma$) and we denote the set of such series by $\mathbb{K}[[z^\Gamma]]_{\mathrm{gb}}$ or simply by $\mathbb{K}[[z^\Gamma]]$.

In [14, Section 2.2] it is shown that $\mathbb{K}[[z^\Gamma]]$ forms a field. It actually forms a valued field for the valuation $v: \mathbb{K}[[z^\Gamma]] \longrightarrow \Gamma \cup \{+\infty\}$ defined by $v(f) := \min \operatorname{supp} f$ (and $v(0) := +\infty$). If $f \in \mathbb{K}[[z^\Gamma]]$ is non-zero, then we call $z^{v(f)}$ the *dominant monomial* of $f$. If $\mathbb{K}$ is an ordered field, then so is $\mathbb{K}[[z^\Gamma]]$, by setting $f > 0 \Longleftrightarrow f_{v(f)} > 0$. In that case, we will write $\mathbb{K}[[z^\Gamma]]^{>}$ for the subset of strictly positive elements.

*Example* 1. We have $f = 1/(1 - z - z^\pi) \in \mathbb{Q}[[z^{\mathbb{R}}]]$ with $\operatorname{supp} f = \mathbb{N} + \pi\,\mathbb{N}$.

*Example* 2. If $x := 1/z$ is a "formal infinitely large variable", then we define $x^{-\Gamma} := z^\Gamma$ and $\mathbb{K}[[x^{-\Gamma}]] := \mathbb{K}[[z^\Gamma]]$.

## 2.2 Iterated series

Given variables $z_1, \ldots, z_n$ and $\alpha = (\alpha_1, \ldots, \alpha_n) \in \Gamma^n$, let $z^\alpha := z_1^{\alpha_1} \cdots z_n^{\alpha_n}$. We totally order $\Gamma^n$ anti-lexicographically via

$$\alpha \leqslant \beta \iff \alpha_n < \beta_n \ \vee \ (\alpha_n = \beta_n \wedge \alpha_{n-1} < \beta_{n-1}) \ \vee \ \cdots \ \vee \ (\alpha_n = \beta_n \wedge \cdots \wedge \alpha_1 = \beta_1)$$

and define the field of grid-based *iterated* series in $z_1,\ldots,z_n$ by $\mathbb{K}[[z_1^\Gamma;\ldots;z_n^\Gamma]] := \mathbb{K}[[z^{\Gamma^n}]]$. We note that $\mathbb{K}[[z_1^\Gamma;\ldots;z_n^\Gamma]] \subseteq \mathbb{K}[[z_1^\Gamma]] \cdots [[z_n^\Gamma]]$ and this inclusion is strict as soon as $n > 1$: e.g., $\sum_{n \in \mathbb{N}} z_1^{-n^2} z_2 \in \mathbb{K}[[z_1^\mathbb{R}]][[z_2^\mathbb{R}]] \setminus \mathbb{K}[[z_1^\mathbb{R}; z_2^\mathbb{R}]]$. Note also that $f = 1/(z_1 + z_2) \in \mathbb{K}[[z_1^\mathbb{R}; z_2^\mathbb{R}]]$ with $\operatorname{supp} f = \mathbb{N}(-1,1) + (-1,0)$.

We will sometimes consider series $f = \sum_{\alpha_1,\ldots,\alpha_n} f_{\alpha_1,\ldots,\alpha_n} z_1^{\alpha_1} \cdots z_n^{\alpha_n}$ in $\mathbb{K}[[z_1^\Gamma;\ldots;z_n^\Gamma]]$ jointly with respect to all variables $z_1, \ldots, z_n$ and write $v_z(f) \in \Gamma^n \cup \{+\infty\}$ for its valuation. On other occasions, we expand $f = \sum_{\alpha_n \in \Gamma} f_{\alpha_n} z_n^{\alpha_n}$ as a series in $z_n$ with coefficients $f_{\alpha_n} \in \mathbb{K}[[z_1^\Gamma;\ldots;z_{n-1}^\Gamma]]$ and write $v_{z_n}(f) \in \Gamma \cup \{+\infty\}$ for its valuation in $z_n$.

## 2.3 Asymptotic relations and the canonical decomposition of a series

Given $f, g \in \mathbb{K}[[z^\Gamma]]$, we will use the following traditional asymptotic notation:

$$\begin{aligned} f = O(g) &\iff f \preccurlyeq g \iff v(f) \geqslant v(g) \\ f = o(g) &\iff f \prec g \iff v(f) > v(g) \\ & \phantom{\iff} f \asymp g \iff v(f) = v(g) \\ & \phantom{\iff} f \sim g \iff v(f-g) > v(g). \end{aligned}$$

For elements $\gamma, \eta \in \Gamma \cup \{+\infty\}$ in the value group (or infinity), we also write $\gamma = O(\eta)$ if $|\gamma| \leqslant n|\eta|$ for some integer $n \geqslant 1$ and $\gamma = o(\eta)$ if $n|\gamma| < |\eta|$ for all integers $n \geqslant 1$. The following asymptotic relations will also be useful:

$$\begin{aligned} f \preccurlyeq\!\!\!\preccurlyeq g &\iff v(f) = O(v(g)) \\ f \prec\!\!\prec g &\iff v(f) = o(v(g)). \end{aligned}$$

Any grid-based series $f \in \mathbb{K}[[z^\Gamma]]$ admits a unique *canonical decomposition*

$$f = f_\succ + f_\asymp + f_\prec,$$

where $f_\succ := \sum_{\alpha < 0} f_\alpha z^\alpha$, $f_\asymp := f_0$, and $f_\prec := \sum_{\alpha > 0} f_\alpha z^\alpha$. We define $\mathbb{K}[[z^\Gamma]]_\succ := \{f \in \mathbb{K}[[z^\Gamma]] : f = f_\succ\}$ and $\mathbb{K}[[z^\Gamma]]_\prec := \{f \in \mathbb{K}[[z^\Gamma]] : f = f_\prec\}$.

## 2.4 Computable series

A field $\mathbb{K}$ is said to be *effective* if we have algorithms for the field operations and zero testing. An ordered field is effective if we also have an algorithm for the ordering. Assume that $\mathbb{K}$ is effective and that $\mathscr{R}$ is an effective ordered subfield of $\mathbb{R}$.

We recursively define a *lazy power series* with coefficients in $\mathbb{K}$ and exponents in $\mathscr{R}$ as an algorithm $f$ that takes no input and that either produces zero or a pair $(c, \alpha) =: f^\sharp$ with $c \in \mathbb{K}$ and $\alpha \in \mathscr{R}$, as well as another lazy power series $f^\flat$. The pair $(c, \alpha)$ actually represents a term $c z^\alpha$ and we will use this latter notation in the sequel. Writing $c_0 z^{\alpha_0} := f^\sharp$, $c_1 z^{\alpha_1} := (f^\flat)^\sharp$, $c_2 z^{\alpha_2} := ((f^\flat)^\flat)^\sharp$, ..., where the sequence stops whenever $((f^\flat)\cdots)^\flat$ produces zero, we allow coefficients $c_i$ to be zero, but we require that $\{\alpha_0, \alpha_1, \alpha_2, \ldots\}$ is a grid-based subset of $\mathscr{R}$. We may thus regard $f$ as a grid-based series $f = c_0 z^{\alpha_0} + c_1 z^{\alpha_1} + c_2 z^{\alpha_2} + \cdots \in \mathbb{K}[[z^\mathscr{R}]]$. Conversely, a series $f \in \mathbb{K}[[z^\mathscr{R}]]$ is said to be *computable* if it can be computed as a lazy power series. We write $\mathbb{K}[[z^\mathscr{R}]]^{\mathrm{com}}$ for the set of such series.

In fact, $\mathbb{K}[[z^{\mathscr{R}}]]^{\text{com}}$ forms a field and the lazy approach allows us to implement the ring operations in an elegant way as follows: given $(c,\lambda)\in\mathbb{K}\times\mathscr{R}$ and non-zero $f,g\in\mathbb{K}[[z^{\mathscr{R}}]]^{\text{com}}$ with $f^{\sharp}=a z^{\alpha}$ and $g^{\sharp}=b z^{\beta}$, we set

$$\begin{aligned}
((f\pm g)^{\sharp},(f\pm g)^{\flat}) &:= \begin{cases} ((a\pm b)\,z^{\alpha},f^{\flat}\pm g^{\flat}) & \text{if } \alpha=\beta\\ (a z^{\alpha},f^{\flat}\pm g) & \text{if } \alpha<\beta\\ (b z^{\beta},f\pm g^{\flat}) & \text{if } \alpha>\beta\end{cases}\\
(((c z^{\lambda})\,f)^{\sharp},((c z^{\lambda})\,f)^{\flat}) &:= (c a z^{\alpha+\lambda},(c z^{\lambda})\,f^{\flat})\\
((fg)^{\sharp},(fg)^{\flat}) &:= (a b z^{\alpha+\beta},(a z^{\alpha})\,g^{\flat}+f^{\flat}g)
\end{aligned}$$

We also take $f\pm 0:=f$, $0-f:=(-1)\,f$, etc. Assuming that $v(g)>0$, we invert $1-g$ as follows

$$\left(\left(\tfrac{1}{1-g}\right)^{\sharp},\left(\tfrac{1}{1-g}\right)^{\flat}\right) = \left(1,g\tfrac{1}{1-g}\right).$$

Since any non-zero $f\in\mathbb{K}[[z^{\mathscr{R}}]]^{\text{com}}$ can be written as $f=c z^{\alpha}(1-g)$ with $c\in\mathbb{K}^{\neq}$, $\alpha\in\mathscr{R}$, and $g\in\mathbb{K}[[z^{\mathscr{R}}]]^{\text{com}}$ with $v(g)>0$, we may thus compute $f^{-1}$ as $c^{-1}z^{-\alpha}\frac{1}{1-g}$.

The lazy approach is very convenient as long as we only need a modest number of terms. For high order expansions, the relaxed (or online) approach is more efficient [13].

## 2.5 Computable iterated series

Assume now that $\mathbb{K}$ is an effective field and that $\mathscr{R}$ is an effective ordered subfield of $\mathbb{R}$. In the case of iterated series in $\mathbb{K}[[z_1^{\mathscr{R}};\ldots;z_n^{\mathscr{R}}]]=\mathbb{K}[[z^{\mathscr{R}^n}]]$, the lazy approach raises the problem of infinite cancellations: assume that we wish to subtract

$$\frac{1}{1-z_1-z_2}-\frac{1}{1-z_1} \tag{2}$$

using the lazy approach. Then the successive terms of the result are $(1-1)\,z_1^0$, $(1-1)\,z_1^1$, $(1-1)\,z_1^2$, .... Due to this infinite cancellation, we never reach the first term $z_2$ of the result.

In order to circumvent this difficulty, assume that we are given an effective subfield $\mathscr{F}$ of $\mathbb{K}[[z_1^{\mathscr{R}};\ldots;z_n^{\mathscr{R}}]]$ such that $\mathbb{K}\cup\{z^{\alpha}:\alpha\in\mathscr{R}^n\}\subseteq\mathscr{F}$. We will call $\mathscr{F}$ an *ambient field* if for any $f\in\mathscr{F}\cap\mathbb{K}[[z_1^{\mathscr{R}};\ldots;z_k^{\mathscr{R}}]]$ with $1\leqslant k\leqslant n$, when regarding $f=\sum_{\alpha\in\mathscr{R}}f_{\alpha}z_k^{\alpha}$ as a series in $\mathbb{K}[[z_1^{\mathscr{R}};\ldots;z_{k-1}^{\mathscr{R}}]][[z_k^{\mathscr{R}}]]$, we have $f_{\alpha}\in\mathscr{F}$ for all $\alpha\in\mathscr{R}$ and $f\in\mathscr{F}[[z_k^{\mathscr{R}}]]^{\text{com}}$.

In the example (2) we may then take $\mathscr{F}:=\mathbb{K}(z_1,z_2)$, after which expansion with respect to $z_2$ yields

$$\frac{1}{1-z_1-z_2}-\frac{1}{1-z_1} = \left(\frac{1}{1-z_1}-\frac{1}{1-z_1}\right)+\frac{z_2}{(1-z_1)^2}+\frac{z_2^2}{(1-z_1)^3}+\cdots$$

and the coefficients $\frac{1}{1-z_1}-\frac{1}{1-z_1}$, $\frac{1}{(1-z_1)^2}$, $\frac{1}{(1-z_1)^3}$, ... are all in $\mathscr{F}$. In particular, we may detect the infinite cancellation in the first term using the zero test in $\mathscr{F}$.

## 2.6 Beyond grid-based series

In this paper, we adopted the framework of grid-based series in order to use some of the results from [14]. However, the results from this book and the present paper can be adapted *mutatis mutandis* to so called *steady series*.

We say that a subset $S$ of $\mathbb{R}$ is *steady* if $S$ is either finite or $S=\{\alpha_0,\alpha_1,\ldots\}$ with $\alpha_0<\alpha_1<\cdots$ and $\lim_{i\to\infty}\alpha_i=+\infty$. It was shown by Levi-Civita [21] that the set $\mathbb{K}[[z^{\mathbb{R}}]]_{\text{st}}$ of series $f=\sum_{\alpha\in\mathbb{R}}f_{\alpha}z^{\alpha}$ with steady support forms a field. An example of such a steady series $f$ that is not grid-based is $f=\zeta(-\log z)=\sum_{n\geqslant 1}z^{\log n}$.

The field of *iterated steady series* is simply $\mathbb{K}[[z_1^{\mathbb{R}};\ldots;z_n^{\mathbb{R}}]]_{\mathrm{st}} := \mathbb{K}[[z_1^{\mathbb{R}}]]_{\mathrm{st}} \cdots [[z_n^{\mathbb{R}}]]_{\mathrm{st}}$. In fact, the framework of steady series would have been slightly better for the present paper, since it is the most natural setting for lazy power series expansions. Furthermore, we have seen that a series in $\mathbb{K}[[z_1^{\mathbb{R}}]]_{\mathrm{gb}} \cdots [[z_n^{\mathbb{R}}]]_{\mathrm{gb}}$ is not necessarily grid-based, so extra efforts are sometimes required to prove this, whenever this indeed is the case.

Even more generally, Hahn showed that the set $\mathbb{K}[[z^{\Gamma}]]_{\mathrm{wo}}$ of series $f = \sum_{\alpha \in \Gamma} f_\alpha z^\alpha$ with well-ordered support also forms a field. In this setting, we naturally have $\mathbb{K}[[z^{\mathbb{R}^n}]]_{\mathrm{wo}} = \mathbb{K}[[z_1^{\mathbb{R}}]]_{\mathrm{wo}} \cdots [[z_n^{\mathbb{R}}]]_{\mathrm{wo}}$. An example of a well-ordered series that is not steady is $f := z^{-1} + z^{-1/2} + z^{-1/4} + \cdots$, which is a natural solution of the equation $f(z) = z^{-1} + f(z^{1/2})$. However, this kind of series is more problematic from an algorithmic point of view, since the order type of the support of $g(z) := f(z)/(1-z)$ is $\omega^2$. In particular, expressions like $g(z) - f(z)$ give rise to infinite cancellations as in (2), but with no easy fix; see also [9].

# 3 Transseries

## 3.1 Transbases

Let $x$ be a formal variable that we think of as being infinitely large. Given $\ell \in \mathbb{N}$, let $\log_\ell := \log \circ \overset{\ell\times}{\cdots} \circ \log$ denote the $\ell$-fold iterated logarithm. A *transbasis* is a tuple $\mathfrak{B} = (\mathfrak{b}_1, \ldots, \mathfrak{b}_n)$ with the following properties:

**TB1.** $\mathfrak{b}_1 = \log_\ell x$ for some $\ell \in \mathbb{N}$.

**TB2.** $\mathfrak{b}_i \in \exp \mathbb{R}[[\mathfrak{b}_1^{-\mathbb{R}};\ldots;\mathfrak{b}_{i-1}^{-\mathbb{R}}]]_{>}^{>}$ for $i = 2, \ldots, n$.

**TB3.** $\log \mathfrak{b}_2 \prec \cdots \prec \log \mathfrak{b}_n$.

In **TB2**, we understand that $\exp \mathbb{R}[[\mathfrak{b}_1^{-\mathbb{R}};\ldots;\mathfrak{b}_{i-1}^{-\mathbb{R}}]]_{>}^{>}$ consists of the multiplicative group of formal exponentials $\mathrm{e}^{\varphi}$ with $\varphi \in \mathbb{R}[[\mathfrak{b}_1^{-\mathbb{R}};\ldots;\mathfrak{b}_{i-1}^{-\mathbb{R}}]]_{\succ}$ and $\varphi > 0$. Accordingly, in **TB3**, we have written $\log \mathfrak{b}_i$ for the series $\varphi_i$ with $\mathfrak{b}_i = \mathrm{e}^{\varphi_i}$.

We may always insert further iterated logarithms into $\mathfrak{B}$ whenever needed. More precisely, the tuple $\hat{\mathfrak{B}} := (\mathfrak{b}_0, \mathfrak{b}_1, \ldots, \mathfrak{b}_n)$ with $\mathfrak{b}_0 := \log_{\ell+1} x$ is again a transbasis. In the extension $\mathbb{R}[[\mathfrak{b}_0^{-\mathbb{R}};\mathfrak{b}_1^{-\mathbb{R}};\ldots;\mathfrak{b}_{n-1}^{-\mathbb{R}}]]_{>}^{>}$, **TB3** can be strengthened to

$$1 \prec \log \mathfrak{b}_1 \prec \cdots \prec \log \mathfrak{b}_n. \tag{3}$$

*Example* 3. The tuple $\mathfrak{B} = (\mathfrak{b}_1, \mathfrak{b}_2, \mathfrak{b}_3, \mathfrak{b}_4) := (\log x, x, x^x, \mathrm{e}^{x^3+x^2+x})$ forms a transbasis, since $x = \mathrm{e}^{\log x}$, $x^x = \mathrm{e}^{x \log x}$ with $x \log x \in \mathbb{R}[[(\log x)^{-\mathbb{R}}; x^{-\mathbb{R}}]]_{>}^{>}$, and $x^3 + x^2 + x \in \mathbb{R}[[(\log x)^{-\mathbb{R}}; x^{-\mathbb{R}}; x^{-x\mathbb{R}}]]_{>}^{>}$.

*Remark* 4. Let $N \geqslant 1$ be a positive integer. Then

$$\exp \frac{x^N}{1-x^{-1}} = \mathrm{e}^{x^N + \cdots + x} \mathrm{e}^{\frac{1}{1-x^{-1}}} \in \mathbb{R}\left[\left[x^{-\mathbb{R}}; \mathrm{e}^{-(x^N + x^{N-1} + \cdots + x)\mathbb{R}}\right]\right],$$

where $(x, \mathrm{e}^{x^N + x^{N-1} + \cdots + x})$ is a transbasis. If $N$ is large, then the expression $x^N + \cdots + x$ becomes very large, which is not convenient. There are various other types of transbasis, for which $\exp \frac{x^N}{1-x^{-1}}$ may directly be included in the transbasis instead of $\mathrm{e}^{x^N + x^{N-1} + \cdots + x}$. In this paper, we will ignore such "optimizations" and refer to [14, Section 4.4] for more details on alternative definitions.

### 3.2 Grid-based transseries

A *grid-based transseries* is an element of $\mathbb{R}[[\mathfrak{b}_1^{-\mathbb{R}};\ldots;\mathfrak{b}_n^{-\mathbb{R}}]]$ for some transbasis $\mathfrak{B}=(\mathfrak{b}_1,\ldots,\mathfrak{b}_n)$. It turns out that the grid-based transseries form a field $\mathbb{T}$ (modulo natural identifications when varying $\mathfrak{B}$). This is not directly obvious from our "definition", which depends on the underlying transbasis $\mathfrak{B}$. Usually, one first defines the field of transseries $\mathbb{T}$ in a more conceptual manner and then proves that any transseries can be expanded with respect to a transbasis: see [14, Chapter 4 and Section 4.4]. However, in this paper, we will always manipulate transseries via transbasis, so our more computational "definition" will be more direct and convenient.

Grid-based transseries were first considered by Écalle in [7]. For constructions of fields of transseries with well-ordered support, we refer to [1, Appendix A] or [4, 10].

### 3.3 Differentiation of transseries

Consider a transbasis $\mathfrak{B}=(\mathfrak{b}_1,\ldots,\mathfrak{b}_n)$ with $\mathfrak{b}_1=\log_\ell x$. Then we have the natural derivation $\delta_1:=x\log x\cdots\log_\ell x\frac{\partial}{\partial x}$ on $\mathbb{R}[[\mathfrak{b}_1^{-\mathbb{R}}]]$, with $\delta_1 f=-\sum_{\alpha\in\mathbb{R}}\alpha f_\alpha\mathfrak{b}_1^{-\alpha}$ for all $f\in\mathbb{R}[[\mathfrak{b}_1^{-\mathbb{R}}]]$. This derivation extends by induction on $n$ to $\mathbb{B}:=\mathbb{R}[[\mathfrak{b}_1^{-\mathbb{R}};\ldots;\mathfrak{b}_n^{-\mathbb{R}}]]$: assuming that we defined $\delta_1$ on $\mathbb{R}[[\mathfrak{b}_1^{-\mathbb{R}};\ldots;\mathfrak{b}_{n-1}^{-\mathbb{R}}]]$, we can in particular compute $\delta_1\mathfrak{b}_n/\mathfrak{b}_n:=\delta_1\log\mathfrak{b}_n\in\mathbb{R}[[\mathfrak{b}_1^{-\mathbb{R}};\ldots;\mathfrak{b}_{n-1}^{-\mathbb{R}}]]$. Now given $f=\sum_{\alpha\in\mathbb{R}}f_\alpha\mathfrak{b}_n^{-\alpha}\in\mathbb{R}[[\mathfrak{b}_1^{-\mathbb{R}};\ldots;\mathfrak{b}_{n-1}^{-\mathbb{R}}]][[\mathfrak{b}_n^{-\mathbb{R}}]]$, we take

$$\delta_1 f \;:=\; \sum_{\alpha\in\mathbb{R}}\left(\delta_1 f_\alpha-\alpha\frac{\delta_1\mathfrak{b}_n}{\mathfrak{b}_n}f_\alpha\right)\mathfrak{b}_n^{-\alpha}\;\in\;\mathbb{R}[[\mathfrak{b}_1^{-\mathbb{R}};\ldots;\mathfrak{b}_{n-1}^{-\mathbb{R}}]][[\mathfrak{b}_n^{-\mathbb{R}}]].$$

If $f\in\mathbb{R}[[\mathfrak{b}_1^{-\mathbb{R}};\ldots;\mathfrak{b}_n^{-\mathbb{R}}]]$, then it can be shown that $\delta_1 f\in\mathbb{R}[[\mathfrak{b}_1^{-\mathbb{R}};\ldots;\mathfrak{b}_n^{-\mathbb{R}}]]$. This is due to the fact that, for a suitable notion of infinite summation, we have

$$\delta_1 f \;=\; -\sum_{\alpha\in\mathbb{R}^n}f_\alpha\left(\alpha_1\frac{\delta_1\mathfrak{b}_1}{\mathfrak{b}_1}+\cdots+\alpha_n\frac{\delta_1\mathfrak{b}_n}{\mathfrak{b}_n}\right)\mathfrak{b}_1^{-\alpha_1}\cdots\mathfrak{b}_n^{-\alpha_n}, \tag{4}$$

with $\operatorname{supp}\delta_1 f\subseteq\operatorname{supp}f+\operatorname{supp}\frac{\delta_1\mathfrak{b}_1}{\mathfrak{b}_1}+\cdots+\operatorname{supp}\frac{\delta_1\mathfrak{b}_n}{\mathfrak{b}_n}$. For details, see [14, Sections 2.4 and 5.1].

Assuming that $\log_\ell x,\log_{\ell-1}x,\ldots,x$ all belong to $\mathfrak{B}$, it follows that $\mathbb{B}$ is also closed under the usual differentiation with respect to $x$. In addition, for $i=2,\ldots,n$, we have the derivation $\delta_i:=(\mathfrak{b}_i/\delta_1\mathfrak{b}_i)\,\delta_1$ with $\delta_i\mathfrak{b}_i=\mathfrak{b}_i$. Given $f,g\in\mathbb{B}$, it is shown in [14, Section 5.1] that

$$f\prec g\wedge g\not\asymp 1\;\Longrightarrow\;\delta_i f\prec\delta_i g. \tag{5}$$

Applying this to Equation (3), we get

$$\frac{\delta_i\mathfrak{b}_1}{\mathfrak{b}_1}\prec\cdots\prec\frac{\delta_i\mathfrak{b}_n}{\mathfrak{b}_n}. \tag{6}$$

Given $f\in\mathbb{R}[[\mathfrak{b}_1^{-\mathbb{R}};\ldots;\mathfrak{b}_n^{-\mathbb{R}}]]$, the formula (4) implies $v_n(\delta_i f)=v_n(\delta_1 f)\geqslant v_n(f)$, since $v_n(\mathfrak{b}_i/\delta_1\mathfrak{b}_i)=0$. For any $r\in\mathbb{N}$, it follows that $v_n(\delta_i^r f)\geqslant v_n(f)$. If $f$ has dominant monomial $\mathfrak{b}_1^{-\alpha_1}\cdots\mathfrak{b}_m^{-\alpha_m}$ with $\alpha_m>0$, this also yields $v_m(\delta_i f)>0$, whence $\delta_i f\prec 1$. For all $r,k\in\mathbb{N}$, we thus obtain

$$f\prec 1\;\Longrightarrow\;(\delta_i^r f)^k\prec 1. \tag{7}$$

### 3.4 Computable transseries

Let $\mathscr{R}$ be an effective ordered subfield of $\mathbb{R}$ and let $\mathscr{F} \subseteq \mathscr{R}[[\mathfrak{b}_1^{-\mathscr{R}};\ldots;\mathfrak{b}_n^{-\mathscr{R}}]]$ be an ambient field. We call $\mathscr{F}$ a *differential ambient field* if $\log \mathfrak{b}_2,\ldots,\log \mathfrak{b}_n \in \mathscr{F}$ and $\mathscr{F}$ is effectively closed under $\delta_1$. Differentiation can be implemented in a lazy manner: given a non-zero $f \in \mathscr{F} \cap \mathscr{R}[[\mathfrak{b}_1^{-\mathscr{R}};\ldots;\mathfrak{b}_k^{-\mathscr{R}}]]$ with $f^\sharp = a\, \mathfrak{b}_k^{-\alpha}$ (when regarded as a lazy series in $\mathfrak{b}_k^{-1}$), we may take

$$((\delta_k f)^\sharp, (\delta_k f)^\flat) = ((\delta_k a - \alpha a)\, \mathfrak{b}_k^{-\alpha}, \delta_k f^\flat),$$

where $\delta_k a = 0$ if $k = 1$ and $\delta_k a$ can be expanded recursively as $(\mathfrak{b}_k / \delta_{k-1} \mathfrak{b}_k)\, \delta_{k-1} a$ with respect to $\mathfrak{b}_{k-1}^{-1}$ if $k > 1$.

We already noted that $\hat{\mathfrak{B}} := (\mathfrak{b}_0, \mathfrak{b}_1, \ldots, \mathfrak{b}_n)$ is a transbasis for $\mathfrak{b}_0 := \log \mathfrak{b}_1$. Moreover, $\hat{\mathscr{F}} := \mathscr{F}(\mathfrak{b}_0^{-\mathscr{R}})$ is again a differential ambient field for $\delta_0 := \mathfrak{b}_0 \delta_1$. Indeed, $\mathscr{F} \cup \mathfrak{b}_0^{-\mathscr{R}} \subseteq \hat{\mathscr{F}}$ and the lazy algorithms for the field operations allow us to compute the iterated coefficients of any transseries in the field generated by $\mathscr{F} \cup \mathfrak{b}_0^{-\mathscr{R}}$.

# 4 Linear differential equations

For the rest of this paper, let $\mathscr{R}$ be an effective ordered subfield of $\mathbb{R}$, let $\mathfrak{B} = (\mathfrak{b}_1, \ldots, \mathfrak{b}_n)$ be a transbasis, and let $\mathscr{F} \subseteq \mathscr{R}[[\mathfrak{b}_1^{-\mathscr{R}};\ldots;\mathfrak{b}_n^{-\mathscr{R}}]] \subseteq \mathbb{B} := \mathbb{R}[[\mathfrak{b}_1^{-\mathbb{R}};\ldots;\mathfrak{b}_n^{-\mathbb{R}}]]$ be a differential ambient field. We denote by $\mathbb{T}$ the field of grid-based transseries.

### 4.1 Linear differential equations over transseries

Consider a linear differential operator $L = L_r \delta_1^r + \cdots + L_0 \in \mathbb{B}[\delta_1]$ with $L_r \neq 0$. We define $v(L) := \min(v(L_0), \ldots, v(L_r))$ and $D_L := (L_r)_{v(L)} \delta_1^r + \cdots + (L_0)_{v(L)} \in \mathbb{R}[\delta_1]$.

Given $\alpha \in \mathbb{R}^n$ and $\mathfrak{b}^{-\alpha} := \mathfrak{b}_1^{-\alpha_1} \cdots \mathfrak{b}_n^{-\alpha_n}$, we write $L_{\times \mathfrak{b}^{-\alpha}}$ for the unique differential operator in $\mathbb{B}[\delta_1]$ with $L_{\times \mathfrak{b}^{-\alpha}}(f) = L(\mathfrak{b}^{-\alpha} f)$ for all $f \in \mathbb{B}$, and we let $L_{\ltimes \mathfrak{b}^{-\alpha}} := \mathfrak{b}^\alpha L_{\times \mathfrak{b}^{-\alpha}}$. We also denote by $\nu_\alpha$ or $\nu_{L,\alpha}$ the integer such that $L_{\times \mathfrak{b}^{-\alpha}} = (L_{\times \mathfrak{b}^{-\alpha}})_r \delta_1^r + \cdots + (L_{\times \mathfrak{b}^{-\alpha}})_{\nu_\alpha} \delta_1^{\nu_\alpha}$ and $(L_{\times \mathfrak{b}^{-\alpha}})_{\nu_\alpha} \neq 0$.

**Theorem 5.**

*a*) *All transseries solutions of the equation* $Lf = 0$ *in* $\mathbb{T}$ *are actually in* $\mathbb{B}[\log \mathfrak{b}_1]$.

*b*) *If* $f \in \mathbb{B}[\log \mathfrak{b}_1]$ *is a solution of* $Lf = 0$ *with dominant monomial* $(\log \mathfrak{b}_1)^i\, \mathfrak{b}^{-\alpha}$, *then we have* $i < \nu_\alpha$. *Conversely, any* $\alpha \in \mathbb{R}^n$ *and* $i < \nu_\alpha$ *give rise to such a solution.*

*c*) *Given* $g \in \mathbb{B}[\log \mathfrak{b}_1]$, *all solutions of* $Lf = g$ *in* $\mathbb{T}$ *are actually in* $\mathbb{B}[\log \mathfrak{b}_1]$.

*d*) *Moreover, there exists a unique solution to* $Lf = g$ *in* $\mathbb{B}[\log \mathfrak{b}_1]$ *with the property that, for any* $\alpha \in \mathbb{R}^n$ *and* $i < \nu_\alpha$, *the coefficient of* $(\log \mathfrak{b}_1)^i\, \mathfrak{b}^{-\alpha}$ *in* $f$ *vanishes.*

*Proof.* The statements are rephrasings of [14, Theorem 7.17] and its corollaries. □

In $(d)$, the unique solution $f$ is called the *distinguished solution* of $Lf = g$ and we denote it by $f := L^{-1} g$. The operator $L^{-1}: \mathbb{B}[\log \mathfrak{b}_1] \longrightarrow \mathbb{B}[\log \mathfrak{b}_1]$ is linear and even preserves infinite summation [14, Section 7.4].

### 4.2 Computing distinguished solutions

Given $L \in \mathscr{F}[\delta_1]$, let $\mathbb{B}_L := \{g \in \mathbb{B} : L^{-1} g \in \mathbb{B}\}$. We say that $\mathscr{F}$ is *effectively linearly closed* if $L^{-1}: \mathscr{F} \cap \mathbb{B}_L \longrightarrow \mathscr{F}$ for every $L \in \mathscr{F}[\delta_1]$ and if we have an algorithm to compute $L^{-1}$. In order to make $\mathscr{F}$ effectively linearly closed, we need to consider sequences of extensions $\mathscr{F} \subseteq \mathscr{F}(\delta_1^{\mathbb{N}} f)$ by distinguished solutions $f$ of linear differential equations. The main difficulty concerns the design of a zero-test on such an extension $\mathscr{F}(\delta_1^{\mathbb{N}} f)$. In this subsection, we will start by showing how to expand distinguished solutions with respect to $\mathfrak{b}_n$, assuming that $\mathscr{H} := \mathscr{F} \cap \mathbb{R}[[\mathfrak{b}_1^{-\mathbb{R}};\ldots;\mathfrak{b}_{n-1}^{-\mathbb{R}}]]$ is effectively linearly closed.

So consider a linear differential operator $L = L_r\,\delta_n^r + \cdots + L_0 \in \mathscr{F}[\delta_n]$ with $L_r \neq 0$. Note that we regard $L$ as an operator in $\delta_n$ instead of $\delta_1$, for convenience. We define $v_n(L) := \min\,(v_n(L_0), \ldots, v_n(L_r))$, where $v_n$ stands for the valuation with respect to $\mathfrak{b}_n^{-1}$.

Now let $g \in \mathscr{F}[\log \mathfrak{b}_1]$. If $g = 0$, then $L^{-1} g = 0$, so assume that $g \neq 0$ and let $g^\sharp =: \varphi\,\mathfrak{b}_n^{-\beta}$ be the first term of its expansion in $\mathfrak{b}_n^{-1}$. In order to compute the expansion of $L^{-1} g$ in $\mathfrak{b}_n^{-1}$, we may assume without loss of generality that $v_n(L) = 0$ and decompose $L = H + T$ with $H \in \mathscr{H}[\delta_n] = \mathscr{H}[\delta_{n-1}]$ and $T \in \mathscr{F}[\delta_n]$ with $v_n(T) > 0$. We may now expand $L^{-1} g$ in a lazy manner as follows:

$$((L^{-1} g)^\sharp, (L^{-1} g)^\flat) \;:=\; \big(\psi\,\mathfrak{b}_n^{-\beta}, L^{-1}\big(g^\flat - T\big(\psi\,\mathfrak{b}_n^{-\beta}\big)\big)\big), \qquad \text{where } \psi := H_{\ltimes \mathfrak{b}_n^{-\beta}}^{-1}\,\varphi. \tag{8}$$

Here we note that $H_{\ltimes \mathfrak{b}_n^{-\beta}} \in \mathscr{H}[\delta_{n-1}]$, whence $\psi \in \mathscr{H}$ using our assumption that $\mathscr{H}$ is effectively linearly closed.

### 4.3 Indicial polynomials

We extend the classical notion of indicial polynomials [18] to linear differential operators with transseries coefficients. Let $L = L_r\,\delta_n^r + \cdots + L_0 \in \mathscr{F}[\delta_n]$ with $L_r \neq 0$ still be a linear differential operator in $\delta_n$. We define the *indicial polynomial* of $L$ by $\mathrm{I}_L := (L_r)_{v(L)}\,(-N)^r + \cdots + (L_0)_{v(L)} \in \mathscr{R}[N]$. (If $n = 1$, then $D_L = \mathrm{I}_L(-\delta_1)$, but this does not hold in general.)

**Proposition 6.** *Let $\alpha \in \mathbb{R}^n$ and $L = L_r\,\delta_n^r + \cdots + L_0 \in \mathscr{F}[\delta_n]$ with $L_r \neq 0$. If $\nu_{L,\alpha} > 0$, then $\alpha_n$ must be a root of the indicial polynomial $\mathrm{I}_L$.*

*Proof.* First note that $L_{\ltimes \mathfrak{b}_n^{-\alpha_n}} = \mathfrak{b}_n^{\alpha_n} L_{\times \mathfrak{b}_n^{-\alpha_n}}$ can be obtained from $L$ by substituting $\delta_n - \alpha_n$ for $\delta_n$. In particular, $\mathrm{I}_{L_{\ltimes \mathfrak{b}_n^{-\alpha_n}}}(N) = \mathrm{I}_L(N + \alpha_n)$. Then writing $L_{\ltimes \mathfrak{b}_n^{-\alpha_n}} = \tilde{L}_r\,\delta_n^r + \cdots + \tilde{L}_0 + o\big(\mathfrak{b}^{v(L_{\ltimes \mathfrak{b}_n^{-\alpha_n}})}\big)$, we have $\tilde{L}_0 = \mathrm{I}_L(\alpha_n)$. Applying (6) to $\delta_n$ gives $\delta_n\,\mathfrak{b}_i / \mathfrak{b}_i \prec \delta_n\,\mathfrak{b}_n / \mathfrak{b}_n = 1$ for $i = 1, \ldots, n-1$. Therefore, $L_{\times \mathfrak{b}^{-\alpha}} = \mathfrak{b}_n^{\alpha_n} \tilde{L}_r \delta_n^r + \cdots + \mathfrak{b}_n^{\alpha_n} \tilde{L}_0 + o(\mathfrak{b}^{v(L_{\times \mathfrak{b}^{-\alpha}})})$. Then rewrite $L_{\times \mathfrak{b}^{-\alpha}} =: \check{L} = \check{L}_r \delta_1^r + \cdots + \check{L}_0$ in $\delta_1$. Applying (6) to $\delta_1$ yields $1 = \delta_1\,\mathfrak{b}_1 / \mathfrak{b}_1 \prec \delta_1\,\mathfrak{b}_n / \mathfrak{b}_n$ and thus $\check{L} = \mathfrak{b}_n^{\alpha_n}\,\tilde{L}_0 + o(\mathfrak{b}^{v(L_{\times \mathfrak{b}^{-\alpha}})})$. Hence once $\nu_{L,\alpha} > 0$, we must have $\mathfrak{b}_n^{\alpha_n} \tilde{L}_0 = \mathfrak{b}_n^{\alpha_n} \mathrm{I}_L(\alpha_n) = 0$. □

Combining Theorem 5 and Proposition 6, if $f_1$ and $f_2$ are two distinct solutions to $L f = g$ in $\mathscr{F}[\log \mathfrak{b}_1]$, then $v_n(f_2 - f_1)$ must be a root of the indicial polynomial $\mathrm{I}_L$.

## 5 Quasi-linear differential equations

### 5.1 Differential equations over transseries

Consider a differential polynomial $P \in \mathbb{B}[\delta_1^{\mathbb{N}} F]$ of order $r$ and total degree $d$ in $F, \ldots, \delta_1^r F$. We may also write $P = \sum_{i_0, \ldots, i_r \in \mathbb{N}} P_{i_0, \ldots, i_r} F^{i_0} \cdots (\delta_1^r F)^{i_r}$ with $P_{i_0, \ldots, i_r} \in \mathbb{B}$. We define $v(P) = \min_{i_0, \ldots, i_r \in \mathbb{N}} v(P_{i_0, \ldots, i_r})$ and $D_P = \sum_{i_0, \ldots, i_r \in \mathbb{N}} (P_{i_0, \ldots, i_r})_{v(P)} F^{i_0} \cdots (\delta_1^r F)^{i_r}$. We may also decompose $P = P_{[0]} + \cdots + P_{[d]}$, where $P_{[i]}$ regroups all homogeneous terms of total degree $i$ in $F, \ldots, \delta_1^r F$, for $i = 0, \ldots, d$.

Given $\varphi \in \mathbb{B}$, we write $P_{+\varphi}$ and $P_{\times\varphi}$ for the unique differential polynomial in $\mathbb{B}[\delta_1^{\mathbb{N}} F]$ with $P_{+\varphi}(f) = P(\varphi + f)$ and $P_{\times\varphi}(f) = P(\varphi f)$ for all $f \in \mathbb{B}$, respectively. From (7), we get

$$\varphi \prec 1 \implies P_{+\varphi} \;=\; P + o(P). \tag{9}$$

In other words, if $\varphi \prec 1$, then $v(P_{+\varphi} - P) > v(P)$, so in particular $D_{P_{+\varphi}} = D_P$. If $P = P_{[d]}$ is homogeneous of degree $d$ and $\varphi \neq 0$, then one also has

$$v_n(P_{\times\varphi}) \;=\; v_n(P) + d\,v_n(\varphi). \tag{10}$$

For details about these elementary definitions and properties, see [14, Sections 8.1, 8.2].

### 5.2 Quasi-linear differential equations

The equation

$$P(f) = 0, \qquad f \prec 1 \tag{11}$$

is said to be *quasi-linear* if $v(P) = v(P_{[1]}) < v(P_{[0]})$. More generally, if $\alpha \in \mathbb{R}^n$, then

$$P(f) = 0, \qquad f \prec \mathfrak{b}^{-\alpha} \tag{12}$$

is said to be *quasi-linear* if $P_{\times \mathfrak{b}^{-\alpha}}(\tilde{f}) = 0$, $\tilde{f} \prec 1$ is quasi-linear [14, Section 8.5]. If $\varepsilon \prec 1$ and (11) is quasi-linear, then so is the equation $P_{+\varepsilon}(\tilde{f}) = 0$, $\tilde{f} \prec 1$, by (9). Consequently, if $\varepsilon \prec \mathfrak{b}^{-\alpha}$ and (12) is quasi-linear, then so is $P_{+\varepsilon}(\tilde{f}) = 0$, $\tilde{f} \prec \mathfrak{b}^{-\alpha}$.

*Example* 7. Let $\mathfrak{B} = (x, \mathrm{e}^x)$, $\mathscr{F} := \mathbb{Q}(x, \mathrm{e}^x)$, and

$$P := \frac{x-1}{\mathrm{e}^{2x}} + \left(\frac{1}{x} - \frac{1}{\mathrm{e}^x} + \frac{1}{x\,\mathrm{e}^x}\right)\delta_1 F + \left(1 - \frac{x}{\mathrm{e}^x}\right) F + \frac{F\,\delta_1 F}{x} + F^2 \in \mathscr{F}[\delta_1^{\mathbb{N}} F].$$

The equation $P(f) = 0$, $f \prec 1$ is quasi-linear, since $v(P) = v(P_{[1]}) = (0,0) < v(P_{[0]}) = (-1,2)$.

Assume that (12) has a solution $f \in \mathbb{B}$ and let $L \in \mathbb{B}[\delta_1]$ be such that $LF = (P_{+f})_{[1]}$. Then the solution $f$ is said to be *distinguished* if $f_\alpha = 0$ for all $\alpha \in \mathbb{R}^n$ with $\nu_{L,\alpha} > 0$.

**Theorem 8.** *Any quasi-linear differential equation* (12) *has a unique distinguished solution in* $\hat{\mathbb{B}} = \mathbb{R}[[\hat{\mathfrak{b}}_1^{-\mathbb{R}}; \ldots; \hat{\mathfrak{b}}_{n+r}^{-\mathbb{R}}]]$*, where* $\hat{\mathfrak{B}} = (\hat{\mathfrak{b}}_1, \ldots, \hat{\mathfrak{b}}_{n+r}) = (\log_r \mathfrak{b}_1, \ldots, \log \mathfrak{b}_1, \mathfrak{b}_1, \ldots, \mathfrak{b}_n)$*. The uniqueness persists when replacing* $\hat{\mathfrak{B}}$ *by an even larger transbasis.*

*Proof.* This is a rephrasing of [14, Theorem 8.21]. □

*Remark* 9. With the above notation, given another solution $\tilde{f}$ of (12) in $\hat{\mathbb{B}}$, the difference $\varepsilon := \tilde{f} - f$ satisfies the quasi-linear equation $P_{+f}(\varepsilon) = 0$, $\varepsilon \prec \mathfrak{b}^{-\alpha}$. Then $\nu_{L,v(\varepsilon)} > 0$ since otherwise $P(\tilde{f}) = P_{+f}(\varepsilon) \sim L\varepsilon \neq 0$. By Proposition 6, the valuation $v_n(\varepsilon)$ must be a root of the indicial polynomial of $L$, when regarding $L$ as an operator in $\delta_n$. This remark even goes through for complex solutions $\tilde{f} \in \hat{\mathbb{B}}[\mathrm{i}]$ of (12).

### 5.3 Computing distinguished solutions

Consider a normalized quasi-linear differential equation (11), where $P \in \mathscr{F}[\delta_1^{\mathbb{N}} F]$ and $v(P) = 0$. Assume that $\mathfrak{B}$ and $\mathscr{F}$ have been extended such that the distinguished solution $f$ of (11) lies in $\mathbb{B}$. The aim of this subsection is to compute the expansion $f = \sum_{\alpha_n \in \mathscr{R}} f_{\alpha_n} \mathfrak{b}_n^{-\alpha_n}$ of $f$ with respect to $\mathfrak{b}_n^{-1}$.

We first decompose $P = H + T$, where $H \in \mathscr{H}[\delta_1^{\mathbb{N}} F]$ and $T \in \mathscr{F}[\delta_1^{\mathbb{N}} F]$ is such that $v_n(T) := \min_{i_0, \ldots, i_r} v_n(T_{i_0, \ldots, i_r}) > 0$. It is readily checked that $f_0 \in \mathbb{R}[[\mathfrak{b}_1^{-\mathbb{R}}; \ldots; \mathfrak{b}_{n-1}^{-\mathbb{R}}]]$ is the distinguished solution of the equation $H(\varphi) = 0$, $\varphi \prec 1$. We will assume that we already know how to compute this solution $f_0$ and that $f_0 \in \mathscr{H}$. As in Section 4.2, we will also assume that $\mathscr{H}$ is effectively linearly closed.

Now consider $\tilde{P} := P_{+f_0}$ and decompose it as $\tilde{P} =: LF - g - R$, where $L \in \mathscr{H}[\delta_n]$, $g \in \mathscr{F}$ with $v_n(g) > 0$ and $R \in \mathscr{F}[\delta_n^{\mathbb{N}} F]$ with $v_n(R_{[1]}) > 0$ and $v_n(R) \geqslant 0$. Note that we now write $L$ and $R$ with respect to $\delta_n$ instead of $\delta_1$, which does not change their valuation with respect to $\mathfrak{b}_n^{-1}$. We already saw in Section 4.2 how to compute $L^{-1}$. This allows us to lazily compute $\tilde{f} = f - f_0$ using the formula

$$\tilde{f} = L^{-1}(g + R(\tilde{f})).$$

The dependency of the right-hand side on $\tilde{f}$ is legit in the lazy expansion paradigm, as long as the coefficient of $\mathfrak{b}_n^{-\alpha}$ in $L^{-1}(g+R(\tilde{f}))$ only depends on coefficients of $\mathfrak{b}_n^{-\beta}$ in $\tilde{f}$ with $\beta<\alpha$. By construction, this is the case here.

*Example* 10. Following from Example 7, let us compute the expansion of the distinguished solution $U$ to the quasi-linear equation $P(f)=0$, $f\prec 1$. Decomposing $P=H+T$, we have $H=x^{-1}\,\delta_1 F+F+x^{-1}F\,\delta_1 F+F^2$. The distinguished solution of $H(f)=0$, $f\prec 1$ is $U_0=0$, so $\tilde{P}=P_{+U_0}=P$ and $v_2(U)>0$. We now write

$$\tilde{P} \;=\; (x-1)\,\mathrm{e}^{-2x}+(1-x\,\mathrm{e}^{-x}+\mathrm{e}^{-x})\,\delta_2 F+(1-x\,\mathrm{e}^{-x})\,F+F\,\delta_2 F+F^2$$

with respect to $\delta_2$ and decompose it as $\tilde{P}=LF-g-R$, where $L=\delta_2+1$, $g=(1-x)\,\mathrm{e}^{-2x}$ and $R=(x\,\mathrm{e}^{-x}-\mathrm{e}^{-x})\,\delta_2 F+x\,\mathrm{e}^{-x}F-F\,\delta_2 F-F^2$.

Then we compute $U$ using $U=L^{-1}(\gamma)$, where $\gamma=g+R(U)$. By (8), we have $v_2(U)=v_2(\gamma)$. Combining $v_2(R(U))>v_2(U)$, we have $v_2(U)=v_2(g)=2$ and $\gamma^\sharp=g^\sharp=\varphi\,\mathrm{e}^{-2x}$ with $\varphi=1-x$. Decompose $L=\tilde{H}+\tilde{T}$ with $\tilde{H}=\delta_2+1$ and $\tilde{T}=0$. Then we have $\psi:=\tilde{H}^{-1}_{\ltimes\mathrm{e}^{-2x}}\,\varphi=(\delta_2-1)^{-1}\,\varphi=x$. Therefore,

$$((L^{-1}\,\gamma)^\sharp,(L^{-1}\,\gamma)^\flat) \;=\; (x\,\mathrm{e}^{-2x},L^{-1}(\tilde{\gamma})), \qquad \text{where } \tilde{\gamma}:=R(U).$$

Hence the first term of $U$ is $x\,\mathrm{e}^{-2x}$. Repeating this process, we obtain

$$((L^{-1}\,\tilde{\gamma})^\sharp,(L^{-1}\,\tilde{\gamma})^\flat) \;=\; \left(\left(\frac{x^2}{2}-x\right)\mathrm{e}^{-3x},L^{-1}(\tilde{\tilde{\gamma}})\right), \qquad \text{where } \tilde{\tilde{\gamma}}:=\tilde{\gamma}^\flat,$$

$$((L^{-1}\,\tilde{\tilde{\gamma}})^\sharp,(L^{-1}\,\tilde{\tilde{\gamma}})^\flat) \;=\; \left(\left(\frac{x^3}{3}-\frac{3x^2}{2}+x\right)\mathrm{e}^{-4x},L^{-1}(\tilde{\tilde{\tilde{\gamma}}})\right), \qquad \text{where } \tilde{\tilde{\tilde{\gamma}}}:=\tilde{\tilde{\gamma}}^\flat,$$

which leads to the expansion

$$U \;=\; x\,\mathrm{e}^{-2x}+\left(\frac{x^2}{2}-x\right)\mathrm{e}^{-3x}+\left(\frac{x^3}{3}-\frac{3x^2}{2}+x\right)\mathrm{e}^{-4x}+\cdots.$$

# 6 General algebraic differential equations

## 6.1 Existence of solutions

We first recall the basic notion of Newton degree and several related properties. Consider the extension

$$\tilde{\mathfrak{B}} \;=\; (\tilde{\mathfrak{b}}_1,\ldots,\tilde{\mathfrak{b}}_{n+s}) \;=\; (\log_s \mathfrak{b}_1,\ldots,\log \mathfrak{b}_1,\mathfrak{b}_1,\ldots,\mathfrak{b}_n)$$

of $\mathfrak{B}$ with $s$ new logarithms. Let

$$\tilde{\mathbb{B}} \;:=\; \mathbb{R}[[\tilde{\mathfrak{b}}_1^{-\mathbb{R}};\ldots;\tilde{\mathfrak{b}}_{n+s}^{-\mathbb{R}}]]$$

with derivations $\tilde{\delta}_i$ with $\tilde{\delta}_i\,\tilde{\mathfrak{b}}_i=\tilde{\mathfrak{b}}_i$ for $i=1,\ldots,n+s$. Then any $P\in\mathbb{B}[\delta_1^{\mathbb{N}}F]$ can be rewritten as an element $\tilde{P}\in\tilde{\mathbb{B}}[\tilde{\delta}_1^{\mathbb{N}}F]$, which also corresponds to the $s$-fold upward shifting from [14, Section 8.2.3]; see also [1, page 293]. By [14, Theorem 8.6] and its subsequent remark, there exist a unique polynomial $Q\in\mathbb{R}[F]$ and integer $\nu\in\mathbb{N}$ such that $D_{\tilde{P}}=Q\cdot(\tilde{\delta}_1 F)^\nu$ for all sufficiently large $s$. We call $N_P=Q\cdot(\delta_1 F)^\nu$ the *differential Newton polynomial* for $P$. For a general monomial $\mathfrak{b}^{-\alpha}$, the *Newton degree* of the asymptotic differential equation

$$P(f) \;=\; 0, \qquad f \;\prec\; \mathfrak{b}^{-\alpha} \tag{13}$$

is defined to be the largest possible degree of $N_{P_{\times \mathfrak{b}^{-\beta}}}$ for all $\beta \in \mathbb{R}^n$ such that $\alpha < \beta$. In fact, the equation (13) is quasi-linear if and only if its Newton degree is one. The following theorem shows that Newton degree plays a crucial role in determining a lower bound on the number of the solutions to (13) in $\mathbb{T}[\mathrm{i}]$.

**Theorem 11.** *[11, Theorem 35] If the asymptotic algebraic differential equation (13) is of Newton degree d, then (13) has at least d solutions in $\tilde{\mathbb{B}}[\mathrm{i}]$ when counting with multiplicities, provided that s is sufficiently large.*

Note that we can always assume that $\mathfrak{b}^{-\alpha} = 1$ in the equation (13) by replacing $P$ with $P_{\times \mathfrak{b}^{-\alpha}}$. In the sequel, we will therefore only consider the equation

$$P(f) = 0, \qquad f \prec 1. \tag{14}$$

We may then use the following simpler criterion for the existence of solutions:

**Theorem 12.** *Let $P \in \mathbb{B}[\delta_1^{\mathbb{N}} F]$ be such that $(D_P)_{[0]} = 0$. Then the asymptotic differential equation (14) has a solution in $\tilde{\mathbb{B}}[\mathrm{i}]$, provided that s is sufficiently large.*

To show Theorem 12, it is sufficient to prove the following lemma.

**Lemma 13.** *Let $P \in \mathbb{B}[\delta_1^{\mathbb{N}} F]$. Then $(D_P)_{[0]} = 0$ if and only if the Newton degree of the equation (14) is non-zero.*

*Proof.* If $(D_P)_{[0]} \neq 0$, then (7) implies that (14) cannot have any solutions in $\mathbb{T}[\mathrm{i}]$. Theorem 11 then implies that the Newton degree of (14) must vanish. Conversely, assume that $(D_P)_{[0]} = 0$ and let $\tilde{P} \in \tilde{\mathbb{B}}[\tilde{\delta}_1^{\mathbb{N}} F]$ be $s$-fold upward shifting of $P$ as above. Let $\mathfrak{d}$ and $\tilde{\mathfrak{d}}$ stand for the dominant monomials of $P$ and $\tilde{P}$, respectively. Since $\delta_1 = \phi \tilde{\delta}_1$ for some invertible $\phi$ with $\phi \lll \mathfrak{b}_1$, one may verify that $\mathfrak{d} / \tilde{\mathfrak{d}} \lll \mathfrak{b}_1$. Since $(D_P)_{[0]} = 0$, we have $P_{[0]} \preccurlyeq \mathfrak{d}\, \mathfrak{b}_1^{-\alpha}$ for some $\alpha > 0$. Consequently, $\tilde{P}_{[0]} = P_{[0]} \preccurlyeq \tilde{\mathfrak{d}}\, \mathfrak{b}_1^{-\alpha/2}$, whence $(D_{\tilde{P}})_{[0]} = 0$. Taking $s$ sufficiently large such that $D_{\tilde{P}} \in \mathbb{R}[F]\, (\tilde{\delta}_1 F)^{\mathbb{N}}$, it follows that $\deg N_{\tilde{P}} = \deg N_P > 0$. □

### 6.2 From general equations to quasi-linear equations

At first sight, it may seem that the mere resolution of quasi-linear differential equations is far off from solving general algebraic differential equations. But in fact, it is key the to the resolution of more general equations, due to the following consequence of [14, Section 8.7]:

**Theorem 14.** *Let $\mathscr{K}$ be a differential subfield of $\mathbb{T}$ for the derivation $\partial / \partial x: f \mapsto f'$, in which any equation $f' = \varphi' f$ or $f' = \varphi' / \varphi$ with $\varphi \in \mathscr{K}$ has a non-zero solution in $\mathscr{K}$. Assume also that every quasi-linear differential equation with coefficients in $\mathscr{K}$ has a solution in $\mathscr{K}$. Then any root in $\mathbb{T}$ of a differential polynomial in $\mathscr{K}[\partial^{\mathbb{N}} F]$ must be in $\mathscr{K}$.*

The equations $f' = \varphi' f$ and $f' = \varphi' / \varphi$ have $c\, \mathrm{e}^{\varphi}$ and $\log \varphi + c$ as their general solutions, so in order to apply the theorem, it is important that we know how to compute exponentials and logarithms of elements in $\mathscr{F}$ and $\mathscr{F}^{>}$, respectively, while extending $\mathscr{F}$ if necessary. For this, we will use the same technique as in [23], but we actually only need to compute exponentials and logarithms up to constant factors. Moreover, due to the second condition in the theorem, we will assume that we know how to solve quasi-linear equations.

*6.2.1 Computing logarithms*

Given $\varphi \in \mathscr{F}$ with $\varphi = c\,\mathfrak{b}_1^{-\alpha_1} \cdots \mathfrak{b}_n^{-\alpha_n}(1+\varepsilon)$ with $c > 0$ and $\varepsilon \prec 1$, we have

$$\log \varphi \;=\; -\alpha_1 \log \mathfrak{b}_1 - \cdots - \alpha_n \log \mathfrak{b}_n + \log c + \log(1+\varepsilon).$$

We have seen at the end of Section 3.4 how to extend $\mathfrak{B}$ with $\log \mathfrak{b}_1$ if $\alpha_1 \neq 0$, after which $-\alpha_1 \log \mathfrak{b}_1 - \cdots - \alpha_n \log \mathfrak{b}_n \in \mathscr{F}$. If $c \neq 1$, then this may require the extension of the constant field $\mathscr{R}$ with $\log c$, but we may always assume $c = 1$ if we just wish to integrate $\varphi'/\varphi$. Finally, the function $g := \log(1+\varepsilon)$ is the distinguished (and actually unique) solution of the quasi-linear differential equation $g' = \varepsilon'/(1+\varepsilon)$, $g \prec 1$.

*6.2.2 Computing exponentials*

Let $\varphi \in \mathscr{F}$. As long as $\varphi \asymp \log \mathfrak{b}_i$ for some $i$, we may write $\varphi = -\alpha_i \log \mathfrak{b}_i + \tilde{\varphi}$ and continue with $\tilde{\varphi}$ in the role of $\varphi$. This leads to a decomposition

$$\varphi \;=\; -\alpha_n \log \mathfrak{b}_n - \cdots - \alpha_{k+1} \log \mathfrak{b}_{k+1} + \psi,$$

where $\alpha_n, \ldots, \alpha_{k+1} \in \mathscr{R}$, $\psi \in \mathscr{F}$, and either $k = 0$ and $\psi \prec \log \mathfrak{b}_1$, or $1 \leqslant k < n$ and $\log \mathfrak{b}_k \prec \psi \prec \log \mathfrak{b}_{k+1}$, or $k = n$ and $\log \mathfrak{b}_n \prec \psi$. If $\mathscr{F} \ni \psi_{\succ} \neq 0$, then similar arguments as in Section 3.4 show that $\hat{\mathfrak{B}} := (\mathfrak{b}_1, \ldots, \mathfrak{b}_{k-1}, \mathrm{e}^{|\psi_{\succ}|}, \mathfrak{b}_k, \ldots, \mathfrak{b}_n)$ is again a transbasis and $\hat{\mathscr{F}} := \mathscr{F}((\mathrm{e}^{|\psi_{\succ}|})^{-\mathscr{R}})$ a differential ambient field that contains $\mathrm{e}^{|\psi_{\succ}|}$. After this extension (if necessary), we have

$$\mathrm{e}^{\varphi} \;=\; \mathrm{e}^{\psi_{\asymp}} (\mathrm{e}^{|\psi_{\succ}|})^{\pm 1} \mathfrak{b}_k^{-\alpha_k} \cdots \mathfrak{b}_n^{-\alpha_n} \mathrm{e}^{\psi_{\prec}}.$$

If $\psi_{\asymp} = 0$, then this may require the extension of the constant field $\mathscr{R}$ with $\mathrm{e}^{\psi_{\asymp}}$, but we may assume $\psi_{\asymp} = 0$ if we just wish to compute a non-zero solution of $f' = \varphi' f$. We have $\mathrm{e}^{\psi_{\prec}} = 1 + g$, where $g$ is the distinguished (and actually unique) solution of the quasi-linear equation $g' = \psi_{\prec}'(1+g)$, $g \prec 1$.

*Example* 15. Following Example 10, we will expand $\mathrm{e}^{\varphi}$ with $\varphi = \mathrm{e}^x U + \mathrm{e}^x - x \in \mathbb{Q}(x, \mathrm{e}^x)$. First decompose $\varphi$ as $\varphi = -x + \psi_{\succ} + \psi_{\prec}$, where $\psi_{\succ} = \mathrm{e}^x$ and $\psi_{\prec} = \mathrm{e}^x U$. Then extend the transbasis $\mathfrak{B}$ to $(x, \mathrm{e}^x, \mathrm{e}^{\psi_{\succ}})$. For $\mathrm{e}^{\psi_{\prec}}$, consider the distinguished solution $V$ of the quasi-linear equation $\Omega(f) = 0$, $f \prec 1$, where

$$\Omega \;:=\; \mathrm{e}^x\left(U + \frac{1}{x}\delta_1 U\right)(F+1) - \frac{1}{x}\delta_1 F \;\in\; \mathscr{F}(U)[\delta_1^{\mathbb{N}} F].$$

Then $\mathrm{e}^{\psi_{\prec}} - 1 = V = 1 + x\mathrm{e}^{-x} + (x^2 - x)\mathrm{e}^{-2x} + \left(x^3 - \frac{5}{2}x^2 + x\right)\mathrm{e}^{-3x} + \cdots$ and $\mathrm{e}^{\varphi} = \mathrm{e}^{-x}\mathrm{e}^{\mathrm{e}^x}(V+1)$.

# 7 The new algorithm for zero testing

## 7.1 Ritt reduction

Consider the admissible ranking $\trianglelefteqslant$ on $\delta_1^{\mathbb{N}} F$ with $\delta_1^j F \trianglelefteqslant \delta_1^{j'} F$ whenever $j \leqslant j'$. The *leader* of a differential polynomial $P \in \mathbb{B}[\delta_1^{\mathbb{N}} F] \setminus \mathbb{B}$ is the highest variable $\delta^j F$ occurring in $P$ when regarding $P$ as a polynomial in $F, \delta_1 F, \ldots$. We will denote it by $\ell_P$. Considering $P$ as a polynomial in $\ell_P$, the leading coefficient $I_P$ is called the *initial* of $P$ and $S_P = \partial P / \partial \ell_P$ its *separant*; we also define $H_P := I_P S_P$. If $P$ has degree $d$ in $\ell_P$, then the pair $\operatorname{rank} P := (\ell_P, d)$ is called the *Ritt rank* of $P$ and such pairs are ordered lexicographically. We understand that $\operatorname{rank} P := -\infty$ for polynomials $P \in \mathbb{B}$.

Given $P, Q_1, \ldots, Q_s \in \mathbb{B}[\delta^{\mathbb{N}} F] \setminus \mathbb{B}$, we say that $P$ is *reducible* with respect to $Q_1, \ldots, Q_s$ if there exists an $i$ such that $\ell_P \in \delta_1^{\mathbb{N} \setminus \{0\}} \ell_{Q_i}$ or $\ell = \ell_P = \ell_{Q_i}$ and $\deg_\ell P \geqslant \deg_\ell Q_s$. The process of *Ritt reduction* provides us with a relation of the form

$$I_{Q_1}^{\alpha_1} \cdots I_{Q_s}^{\alpha_s} S_{Q_1}^{\beta_1} \cdots S_{Q_s}^{\beta_s} P \;=\; \Theta_1 Q_1 + \cdots + \Theta_s Q_s + R,$$

where $\alpha_1, \ldots, \alpha_s, \beta_1, \ldots, \beta_s \in \mathbb{N}$, $\Theta_1, \ldots, \Theta_s \in \mathbb{B}[\delta_1^{\mathbb{N}} F][\delta_1]$, $R \in \mathbb{B}[\delta_1^{\mathbb{N}} F]$ and where $R$ is reduced with respect to $Q_1, \ldots, Q_s$. We will denote $R = P \operatorname{rem} Q = P \operatorname{rem} (Q_1, \ldots, Q_s)$.

## 7.2 Root separation bounds

Assume that we are given a grid-based transseries $f \in \mathbb{B}$ such that $f \prec 1$ and a differential polynomial $P \in \mathbb{B}[\delta_1^{\mathbb{N}} F]$. Let $L_{P,f} \in \mathbb{B}[\delta_1]$ be such that $L_{P,f} F = (P_{+f})_{[1]}$. Note that we have $L_{P,f} \neq 0$ if $S_P(f) \neq 0$. We first present a criterion for the existence of a root of $P$ that is sufficiently close to $f$.

**Proposition 16.** *Let $P \in \mathbb{B}[\delta_1^{\mathbb{N}} F] \setminus \mathbb{B}$ and $f \in \mathbb{B}$ be such that $L_{P,f} \neq 0$ and $f \prec 1$. Let $\sigma > 0$. If $v_n(P(f)) > v_n(L_{P,f}) + \sigma$, then $P$ has a root $\tilde{f}$ in $\mathbb{C}[[(\log_s \mathfrak{b}_1)^{-\mathbb{R}}; \ldots; (\log \mathfrak{b}_1)^{-\mathbb{R}}; \mathfrak{b}_1^{-\mathbb{R}}; \ldots; \mathfrak{b}_n^{-\mathbb{R}}]]$ with $v_n(\tilde{f} - f) > \sigma$, for some $s \in \mathbb{N}$.*

*Proof.* Let $\eta := v_n(P(f)) - (v_n(L_{P,f}) + \sigma) > 0$ and $Q := (P_{+f})_{\times \mathfrak{b}_n^{-\sigma - \eta/2}}$. We have $v_n(P(f)) = v_n((P_{+f})_{[0]})$ and $v_n(L_{P,f}) = v_n((P_{+f})_{[1]})$. Using (10), we now get $v_n(Q_{[0]}) > v_n(Q_{[1]}) \geqslant v_n(Q)$. Hence $(D_Q)_{[0]} = 0$, so $Q$ has a root $\varepsilon$ in $\mathbb{C}[[(\log_s \mathfrak{b}_1)^{-\mathbb{R}}; \ldots; (\log \mathfrak{b}_1)^{-\mathbb{R}}; \mathfrak{b}_1^{-\mathbb{R}}; \ldots; \mathfrak{b}_n^{-\mathbb{R}}]]$ with $\varepsilon \prec 1$ and $s \in \mathbb{N}$, by Theorem 12. Then $\tilde{f} = f + \varepsilon\, \mathfrak{b}_n^{-\sigma - \eta/2}$ is as required. □

Now assume that $P(f) = 0$, $f \prec 1$ is quasi-linear and let $f$ be a solution of it. Let $\mathrm{I}_{P,f}$ be the indicial polynomial of $L_{P,f}$, considered as an operator with respect to $\delta_n$. We define $Z_{P,f}$ to be the largest root of $\mathrm{I}_{P,f}$ in $\mathbb{R}$. If no such root exists, then we set $Z_{P,f} := v_n(f)$.

Let us show now that there always exists a threshold $\sigma \geqslant v_n(f)$ in $\mathbb{R}$ such that for any $s \in \mathbb{N}$ and any $\tilde{f}$ in $\mathbb{C}[[(\log_s \mathfrak{b}_1)^{-\mathbb{R}}; \ldots; (\log \mathfrak{b}_1)^{-\mathbb{R}}; \mathfrak{b}_1^{-\mathbb{R}}; \ldots; \mathfrak{b}_n^{-\mathbb{R}}]]$ with $P(\tilde{f}) = 0$ and $v_n(\tilde{f} - f) > \sigma$, we have $\tilde{f} = f$. The smallest such $\sigma$ will be denoted by $\sigma_{P,f}$ and we call it the *root separation* of $P$ at $f$.

**Proposition 17.** *Let $P \in \mathbb{B}[\delta_1^{\mathbb{N}} F] \setminus \mathbb{B}$ such that $P(f) = 0$, $f \prec 1$ is quasi-linear. Assume that $f \in \mathbb{B}$ is a solution of this equation. Then*

$$\sigma_{P,f} \;\leqslant\; \max(v_n(f), Z_{P,f}).$$

*Proof.* Consider a root $\tilde{f} \in \mathbb{C}[[(\log_s \mathfrak{b}_1)^{-\mathbb{R}}; \ldots; (\log \mathfrak{b}_1)^{-\mathbb{R}}; \mathfrak{b}_1^{-\mathbb{R}}; \ldots; \mathfrak{b}_n^{-\mathbb{R}}]]$ of $P$ with $\tilde{f} \prec 1$ and $\tilde{f} \neq f$. Then Remark 9 implies that $v_n(\tilde{f} - f)$ is a root of $\mathrm{I}_{P,f}$. Consequently $\sigma_{P,f} \leqslant \max(v_n(f), Z_{P,f})$. □

## 7.3 The main algorithm

Consider a quasi-linear differential equation $P(f) = 0$, $f \prec 1$ with $P \in \mathscr{F}[\delta_1^{\mathbb{N}} F]$ and assume that its distinguished solution $f$ belongs to $\mathbb{B}$. Given $Q_1, \ldots, Q_s \in \mathscr{F}[\delta_1^{\mathbb{N}} F]$, we will now present an algorithm to check whether $Q_1, \ldots, Q_s$ simultaneously vanish at $f$. By induction, we suppose that a zero test on $\mathscr{H} := (\mathbb{R}[[\mathfrak{b}_1^{-\mathbb{R}}; \ldots; \mathfrak{b}_{n-1}^{-\mathbb{R}}]] \cap \mathscr{F})(\delta_1^{\mathbb{N}} F)$ has been given. In particular, one can test whether the coefficients of $f$ with respect to $\mathfrak{b}_n^{-1}$ are zero and thus compute the valuation of $f$ in $\mathfrak{b}_n^{-1}$.

**Algorithm ZeroTest**$(Q_1,\ldots,Q_s)$
INPUT: $Q_1,\ldots,Q_s \in \mathscr{F}[\delta_1^{\mathbb{N}} F] \setminus \{0\}$, ordered by non-decreasing Ritt rank
OUTPUT: **true** if $Q_1(f) = \cdots = Q_s(f) = 0$ and **false** otherwise

1 If $Q := Q_1 \in \mathscr{F}$ then return **false**
2 If **ZeroTest**$(I_Q)$ then return **ZeroTest**$(I_Q, Q_1, \ldots, Q_s)$
3 If **ZeroTest**$(S_Q)$ then return **ZeroTest**$(S_Q, Q_1, \ldots, Q_s)$
4 If $\exists J \in \{Q_2, \ldots, Q_s, P\}, J \operatorname{rem} Q \neq 0$ then return **ZeroTest**$(J \operatorname{rem} Q, Q_1, \ldots, Q_s)$
5 Let $\sigma := \max(v_n(f), v_n(L_{P,f}), Z_{P,f}, v_n(I_Q(f)), v_n(S_Q(f)))$
6 Return the result of the test $v_n(Q(f)) > \sigma + v_n(L_{Q,f})$

*Proof.* In steps 2, 3, and 4, the Ritt rank of the first argument of **ZeroTest** always strictly decreases; this proves the termination of the algorithm. Furthermore, if one of the tests in steps 2, 3, or 4 succeeds, then the algorithm is clearly correct. In step 1, note that we assumed that $Q_1 \neq 0$ as an element of $\mathscr{F}[\delta_1^{\mathbb{N}} F]$. So if $Q_1 \in \mathscr{F}$, then $Q_1(f) = Q_1 \neq 0$.

Assume now that we reach step 5. By construction, this means that $H_Q(f) \neq 0$ and $J \operatorname{rem} Q = 0$ for all $J \in \{Q_2, \ldots, Q_s, P\}$. In particular, Ritt reduction of $P$ by $Q$ yields a relation

$$I_Q^j S_Q^k P \;=\; U_0 Q + \cdots + U_r \delta_1^r Q, \tag{15}$$

where $j, k \in \mathbb{N}$ and $U_0, \ldots, U_r \in \mathbb{B}[\delta_1^{\mathbb{N}} F]$. If $v_n(Q(f)) \leqslant \sigma + v_n(L_{Q,f})$, then we clearly have $Q(f) \neq 0$, so assume that $v_n(Q(f)) > \sigma + v_n(L_{Q,f})$. Applying Proposition 16, we obtain a grid-based transseries $\tilde{f} \in \mathbb{C}[[(\log_s \mathfrak{b}_1)^{-\mathbb{R}}; \ldots; (\log \mathfrak{b}_1)^{-\mathbb{R}}; \mathfrak{b}_1^{-\mathbb{R}}; \ldots; \mathfrak{b}_n^{-\mathbb{R}}]]$ such that $Q(\tilde{f}) = 0$ and $v_n(\tilde{f} - f) > \sigma$, for some $s \in \mathbb{N}$. Using (7), it follows that

$$v_n(L_{P,\tilde{f}}) \;=\; v_n(L_{P,f}) \;\leqslant\; \sigma, \tag{16}$$

$$v_n(I_Q(\tilde{f})) \;=\; v_n(I_Q(f)) \;\leqslant\; \sigma \quad \text{and} \quad v_n(S_Q(\tilde{f})) \;=\; v_n(S_Q(f)) \;\leqslant\; \sigma. \tag{17}$$

Now (16) implies $\mathrm{I}_{P,\tilde{f}} = \mathrm{I}_{P,f}$ and thus $Z_{P,\tilde{f}} = Z_{P,f} \leqslant \sigma$. From (17), we in particular get $\mathrm{I}_Q(\tilde{f}) \neq 0$ and $S_Q(\tilde{f}) \neq 0$. Hence evaluating (15) at $\tilde{f}$ yields $P(\tilde{f}) = 0$. Applying Proposition 17 to $\tilde{f}$, we obtain $\sigma_{P,\tilde{f}} \leqslant \sigma$. Since $v_n(\tilde{f} - f) > \sigma$, we conclude that $\tilde{f} = f$. From $Q(f) = 0$, Ritt reduction of $J$ by $Q$ also yields $J(f) = 0$ for all $J \in \{Q_2, \ldots, Q_n\}$. □

**Theorem 18.** *Let $\mathscr{F} \subseteq \mathscr{R}[[\mathfrak{b}_1^{-\mathscr{R}}; \ldots; \mathfrak{b}_n^{-\mathscr{R}}]] \subseteq \mathbb{T}$ be a differential ambient field (which comes with a zero test, by assumption). Let $P \in \mathscr{F}[\delta_1^{\mathbb{N}} F] \setminus \mathscr{F}$ be a differential polynomial. If $P(f) = 0, f \prec 1$ is a quasi-linear equation with a distinguished solution $f \in \mathscr{R}[[\mathfrak{b}_1^{-\mathscr{R}}; \ldots; \mathfrak{b}_n^{-\mathscr{R}}]]$, then $\mathscr{F}(\delta_1^{\mathbb{N}} f) \subseteq \mathscr{R}[[\mathfrak{b}_1^{-\mathscr{R}}; \ldots; \mathfrak{b}_n^{-\mathscr{R}}]]$ is again a differential ambient field.*

*Proof.* Our zero test on $\mathscr{F}[\delta_1^{\mathbb{N}} f]$ trivially extends to the quotient field $\mathscr{F}(\delta_1^{\mathbb{N}} f)$ since a fraction vanishes if and only if its numerator does. We may use the algorithms from Sections 2.4, 3.4, and 5.3 to compute expansions of elements in $\mathscr{F}(\delta_1^{\mathbb{N}} f)$. □

### 7.4 A worked example: the Lambert $W$ function

Following from Examples 7, 10 and 15, consider the distinguished transseries solution $U(x) = U$ of $P(f) = 0, f \prec 1$ and let

$$W(x) \;:=\; \log x \cdot U(\log \log x) + \log x - \log \log x.$$

We will use **ZeroTest** to show that $W(x)$ satisfies

$$W(x)\,\mathrm{e}^{W(x)} \;=\; x,$$

i.e., $W=W(x)$ is a real branch of the *Lambert W function* at infinity [3]. To this end, it is enough to verify that

$$(\mathrm{e}^x U+\mathrm{e}^x-x)\,\mathrm{e}^{\mathrm{e}^x U+\mathrm{e}^x-x} \;=\; \mathrm{e}^{\mathrm{e}^x}. \tag{18}$$

In Example 15, we computed $\mathrm{e}^{\mathrm{e}^x U+\mathrm{e}^x-x}=\mathrm{e}^{-x}\mathrm{e}^{\mathrm{e}^x}(V+1)$. Then the equality (18) is equivalent to $Q(V)=0$, where

$$Q \;:=\; (U+1-x\mathrm{e}^{-x})\,(1+F)-1 \;\in\; \mathscr{F}(U)[\delta_1^{\mathbb{N}} F].$$

Now we apply **ZeroTest** algorithm to $\Omega$, $V$ and $Q$ over $\mathscr{F}$. Since $I_Q=S_Q=U+1-x\mathrm{e}^{-x}$, we have $v_2(I_Q(V))=v_2(S_Q(V))=0$ and thus the algorithm proceeds to step 4. Ritt reduction yields

$$I_Q^2\,\Omega \;=\; -x^{-1} I_Q \delta_1 Q+\mathrm{e}^x P(U)\,Q+\mathrm{e}^x P(U).$$

Since $U$ is a root of $P$, we obtain $\Omega$ rem $Q=0$, leading the algorithm to step 5. Writing $L_{\Omega,V} F=\mathrm{e}^x\,(U+x^{-1}\,\delta_1\,U)\,F-x^{-1}\,\delta_1\,F=\mathrm{e}^x\,(U+\delta_2\,U)\,F-\delta_2\,F$, we have $v_2(L_{\Omega,V})=0$, $\mathrm{I}_{\Omega,V}(N)=N$, and thus $Z_{\Omega,V}=0$. Hence, we take $\sigma:=1$. Since $L_{Q,V}F=I_Q F$, we only need to test whether $v_2(Q(V))>1$ in the final step 6:

$$Q(V) \;=\; (1-x\mathrm{e}^{-x}+O(x\mathrm{e}^{-2x}))\,(1+x\mathrm{e}^{-x}+O(x^2\mathrm{e}^{-2x}))-1 \;=\; O(x^2\mathrm{e}^{-2x}).$$

As a result, we finally obtain that $V$ is a root of $Q$ and thus show that $W$ is a real branch of the Lambert $W$ function.